\documentclass[fleqn,usenatbib]{mnras}

\usepackage{newtxtext,newtxmath}

\usepackage[T1]{fontenc}
\usepackage{ae,aecompl}

\usepackage{graphicx}	
\usepackage{amsmath}	
\usepackage{amssymb}	

\usepackage{graphicx}
\usepackage{amssymb}
\usepackage{subfigure}
\newcommand\msun{{\,M_\odot}}

\newcommand\zsun{{\rm \,Z_\odot}}

\newcommand{\unit}[1]{\ensuremath{\, \mathrm{#1}}}
\usepackage{aecompl}
\usepackage[T1]{fontenc}
\usepackage{threeparttable}
\usepackage{caption}
\usepackage{times}
\usepackage{ifpdf}

\newcommand{\cMpc}{~\mbox{comoving}~\mbox{Mpc}}

\newcommand{\pc}{~\mbox{pc}}

\newcommand{\ckpc}{~\mbox{comoving}~\mbox{kpc}}

\newcommand{\cmci}{~\mbox{cm}^{-3}}

\newcommand{\yr}{~\mbox{yr}}

\title[Signature of the first galaxies]{Signature of the first galaxies in JWST deep field observations}

\author[M. Jeon and V. Bromm]{Myoungwon Jeon$^{1}$\thanks{E-mail:
myjeon@khu.ac.kr} and Volker Bromm$^{2}$\\
$^{1}$Center for Global Converging Humanities, Kyung Hee University, Republic of Korea \\ $^{2}$Department of Astronomy, University of Texas, Austin, TX 78712, USA}

\date{Accepted XXX. Received YYY; in original form ZZZ}

\pubyear{2019}

\begin{document}
\label{firstpage}
\pagerange{\pageref{firstpage}--\pageref{lastpage}}
\maketitle

\begin{abstract}
We examine the assembly process and the observability of a first galaxy ($M_{\rm vir}\approx10^9\msun$ at $z\approx8$) with cosmological zoom-in, hydrodynamic simulations, including the radiative, mechanical, and chemical feedback exerted by the first generations of stars. To assess the detectability of such dwarf systems with the upcoming James Webb Space Telescope {\sc (JWST)}, we construct the spectral energy distribution for the simulated galaxy in a post-processing fashion. We find that while the non-ionizing UV continuum emitted by the simulated galaxy is expected to be below the {\sc JWST} detection limit, the galaxy might be detectable using its nebular emission, specifically in the H$\alpha$ recombination line. This requires that the galaxy experiences an active starburst with a star formation rate of $\dot{M}_{\ast}\gtrsim 0.1 \msun \mbox{\,yr}^{-1}$ at $z\approx9$. Due to the bursty nature of star formation in the first galaxies, the time interval for strong nebular emission is short, less than $2-3$~Myr. The probability of capturing such primordial dwarf galaxies during the observable part of their duty cycle is thus low, resulting in number densities of order one source in a single pointing with MIRI onboard the {\sc JWST}, for very deep exposures. Gravitational lensing, however, will boost their observability beyond this conservative baseline. The first sources of light will thus come firmly within our reach.
\end{abstract}

\begin{keywords}
cosmology: theory -- galaxies: formation -- galaxies: high-redshift -- HII regions --
hydrodynamics -- intergalactic medium -- supernovae: physics.
\end{keywords}



\section{Introduction}

One of the biggest challenges in modern cosmology, observationally and theoretically, is to 
understand the properties of the first galaxies, which formed a few hundred million years after the Big Bang \citep{Bromm2011, Loeb2013, Dayal2018}. Such early galaxies are the basic building blocks of 
the massive systems we see today and it is believed that they provided copious ionizing photons that escaped from host galaxies into the intergalactic medium (IGM), thus initiating the process of reionization (for reviews see, e.g. \citealp{Furlanetto2006}; \citealp{Barkana2007}; \citealp{Robertson2010}). The Hubble Space Telescope (HST), however, is limited to detect sources with absolute UV magnitudes $M_{\rm UV}< -19$, implying that the emission from the first galaxies with virial masses of $M_{\rm vir}\sim10^7-10^9\msun$ at $z>7$ is too faint for detection with current instruments (e.g. \citealp{Pawlik2011a}). Therefore, finding the first galaxies is one of the main goals of upcoming telescopes, such as the {\it James Webb Space Telescope} (JWST) and the next generation of giant, 30-40~meter, ground-based telescopes.
\par
What indeed constitutes a first galaxy (e.g. \citealp{Willman2012}; \citealp{Frebel2015})? Theory has suggested galaxies which were hosted by atomic cooling haloes, inside of which gas cooled via atomic hydrogen lines instead of molecular hydrogen ($\rm H_{\rm 2}$). When the first generation of stars, the so-called Population~III (Pop~III), formed in minihaloes with a mass of $M_{\rm vir}\sim10^6\msun$ at $z>15$, the only available coolant was molecular hydrogen in the absence of metals 
(e.g. \citealp{Haiman1996}; \citealp{Tegmark1997}; \citealp{Bromm2002}; \citealp{Yoshida2003}). As the minihaloes grew in mass, the halo virial temperature eventually reached $T_{\rm vir}\sim10^4 \rm K$, corresponding to $M_{\rm vir}\sim5\times10^7\msun$, above which atomic hydrogen line cooling starts to dominate over $\rm H_{\rm 2}$ cooling. Such systems provide possible first galaxy hosts, because of the more efficient cooling and the deeper potential well to retain photo-heated gas.
\par
Currently, we do not yet have direct observations to guide us towards understanding the nature of the first galaxies. The focus has instead been on theoretical studies to investigate how the first galaxies were assembled, and how the first generation of stars played a role in shaping the conditions for their formation (e.g. \citealp{Maio2010}; \citealp{Wise2012}; \citealp{Johnson2013}; \citealp{Xu2016}; \citealp{Yajima2017}; \citealp{Kimm2017}; \citealp{Wilkins2017}; \citealp{Griffen2018}; \citealp{Ma2018}; \citealp{Jaacks2018a}). Owing to their small size ($r_{\rm vir}=1-3$ kpc) and rather short evolutionary timescales (<1 Gyr), the first galaxies are an ideal target for numerical studies, allowing simulators, in principle, to build up a galaxy in an ab-initio fashion. In previous work, \citet{Jeon2015} performed a zoom-in, radiation hydrodynamical simulation of the assembly process of a $M_{\rm vir}\sim10^8\msun$ halo virializing at $z=10$, taking into account the radiative, mechanical, and chemical feedback exerted by the first stars. The existing ab-initio first galaxy simulations, however, have not yet reached the mass scales required to render them detectable with the upcoming generation of frontier facilities (see below).
\par
Considering the effect of stellar feedback on galaxy formation is crucial, especially for such small systems, because they are highly susceptible to disruption due to their shallow potential well (for a review see, e.g. \citealp{Ciardi2005}). In particular, stellar feedback is responsible for establishing a bursty star formation mode in low-mass galaxies. Specifically, photo-ionization and the corresponding heating by photons emitted from Pop~III stars, followed by subsequent supernova explosions, evacuate the gas inside a host halo, thus suppressing further star formation for a period of time. The gas needs to be replenished, so that the next burst of star formation can be triggered (e.g. \citealp{Hopkins2013, Jeon2014, Ma2018, FaucherGigu2018}). Therefore, the star formation history in such small galaxies  ($M_{\rm vir} < 10^9\msun$) exhibits a rather short duty cycle of a few $\sim 10$ Myr, compared to the near-steady histories in their more massive descendants.
\par
Previous studies have revealed that the first galaxies were already metal-enriched at $z\sim10$, instead of truly pristine systems \citep[e.g.][]{Wise2009,Greif2010}. Metals ejected by Pop~III supernovae facilitated a transition in star formation mode from massive-star dominated Pop~III to more normal, low-mass dominated Population~II (Pop~II), prior to the end of the galaxy assembly process at $z\sim10$ (e.g. \citealp{Jeon2015}). Meanwhile, \citealt{Pawlik2011a} argue that {\sc JWST} is only capable of detecting galaxies, using both non-ionizing continuum and nebular emission, with $M_{\rm vir}>10^9\msun$, corresponding to star formation rates $\dot{M}_{\ast}>0.1\msun \rm yr^{-1}$. 
Thus, the $M_{\rm vir}\approx10^8\msun$ systems simulated in previous, ab-initio work are below the {\sc JWST} detection limit, such that simulations need to target more massive hosts \citep[e.g.][]{Zackrisson2014, Barrow2017, Barrow2018, Pallottini2017, Carilli2017, Moriwaki2018, Arata2018, Katz2018, Katz2019, Ceverino2019}.
\par
An alternative strategy to reach the faintest sources with {\sc JWST} deep-field campaigns is to exploit the magnification boost ($\mu>10$ in extreme cases) from gravitational lensing \citep{Zackrisson2012, Zackrisson2014}. Observationally, \citet{Oesch2016} have pushed the high-$z$ frontier to $z\sim11$, using the HST to detect the galaxy $\rm GN$-$\rm z11$, which is surprisingly bright with a star formation rate of $\dot{M}_{\ast}\sim25\msun \rm yr^{-1}$. By fitting its spectral energy distribution (SED), the corresponding stellar mass is inferred to be $\sim10^9\msun$. However, the expected number density of such luminous galaxies at $z\sim11$ is extremely small ($<0.3$ $\rm deg^{-2}$), rendering the faint, majority population at these epochs still undetected. The recently discovered high-$z$ galaxy MACS1149-JD1 at $z\approx9$ (\citealp{Hashimoto2018}) indicates that gravitational lensing (with $\mu=10$) allows us to directly probe light from the first generation of stars formed at $z\approx15$.
\par
In this study, in order to investigate galaxies that can be observed with upcoming telescopes, we focus on a system ($M_{\rm vir}\sim10^9\msun$) more massive by an order of magnitude compared to our previous work (\citealp{Jeon2015}). Such a galaxy is representative of abundant objects in the early Universe, with a number density of $>100$ deg$^{-2}$ at $z\sim7-10$. We here present the results of a cosmological zoom-in, high-resolution, radiation-hydrodynamics simulation, following the first galaxy assembly process from first principles. We consider realistic descriptions for Pop III/Pop II star formation, photo-ionization and photo-heating from stars, and the mechanical and chemical feedback from supernovae. Our goal is to investigate the nature of $M_{\rm vir}\approx10^9\msun$ galaxies in the early Universe and assess their observability.
\par  
The outline of the paper is as follows. Our numerical methodology is described in Section~2, and the simulation results are presented in Section~3. Finally, we discuss our main conclusions in Section~4. For consistency, all distances are expressed in physical (proper) units unless noted otherwise.
\par

\section{Numerical methodology}
\label{Sec:Metho}
\subsection{Gravity, hydrodynamics, and chemistry}
We have performed radiation hydrodynamic zoom-in simulations using a modified version of the $N$-body/TreePM Smoothed Particle
Hydrodynamics (SPH) code GADGET (\citealp{Springel2001}; \citealp{Springel2005}). In order to generate the multi-scale initial conditions, we use the cosmological initial conditions code MUSIC (\citealp{Hahn2011}). From a preliminary dark matter only simulation using $64^3$ particles in a $L=4.46 \cMpc$ box, we identify a target halo with a virial mass of $\sim10^9\msun$ at $z\approx8$. Consequently, we perform a refinement on particles that exist within a Lagrangian volume encompassing all particles within $10R_{\rm vir}$ at $z\approx8$, where $R_{\rm vir}$ is the virial radius of the halo. The resulting dark matter (DM) and gas masses in the most refined region, approximately $450 \ckpc$ on a side, are $m_{\rm DM} \approx 2500\msun$ and $m_{\rm SPH} \approx 495 \msun$, respectively, equivalent to an effective resolution of $1024^3$. 
\par
We adopt a $\Lambda$CDM cosmology with a matter density parameter of $\Omega_{\rm m}=1-\Omega_{\Lambda}=0.265$, baryon 
density $\Omega_{\rm b}=0.0448$, present-day Hubble expansion rate $H_0 = 71\unit{km\, s^{-1} Mpc^{-1}}$, spectral index $n_{\rm s}=0.963$, and normalization $\sigma_8=0.801$. The softening lengths for both DM and baryonic particles are $\epsilon_{\rm soft}=100$ comoving pc, equal to $1/25$ of the mean DM particle separation, i.e., $\epsilon_{\rm soft}\approx 1.95 h^{-1} \unit{ckpc} (L/25 h^{-1} \unit{cMpc}) (N_{\rm DM}^{1/3}/512)^{-1}$, where the number of DM particles 
is $N_{\rm DM}^{1/3}=1024$. We start our simulations at $z \approx 125$ and follow the cosmic evolution of the DM and gas until the assembly process of the first dwarf galaxy with $M_{\rm vir}\sim10^9\msun$ has been completed at $z\approx8$.
\par
Non-equilibrium rate equations for the primordial chemistry for nine atomic and molecular species ($\rm H, H^{+}, H^{-}, H_{2}, H^{+}_2 , He, He^{+}, He^{++},$ and $\rm e^{-}$), as well as for the three deuterium species $\rm D, D^{+}$, and HD are self-consistently solved every time-step. We consider all relevant primordial cooling processes such as H and He collisional ionization, excitation and recombination cooling, bremsstrahlung, inverse Compton cooling, and collisional excitation cooling of $\rm H_2$ and HD \citep{Johnson2006}. Additional metal cooling, specifically by C, O, and Si species is also taken into account by adapting the metal cooling rates given by \citet{Glover2007}. The chemical network comprises the key species, $\rm C, C^{+}, O, O^{+}, Si, Si^{+}$ and $\rm Si^{++}$. Note that we ignore dust cooling, which tends to be less important in the gas density regime ($n_{\rm H}<10^4\rm cm^{-3}$) explored in this work.

\subsection{Star formation physics}

\subsubsection{Population~III}
Due to the limited numerical resolution, such that individual stars cannot be resolved, we rather allow Pop~III stars to form as a single star cluster. For Pop~III stars, we assume a top-heavy initial mass function (IMF) over the mass range of $[1-150]\msun$ with a functional form of 
\begin{equation}
\phi = \frac{dN}{d \ln M} \propto M^{-1.3} \exp{\left[ -\left(\frac{M_{\rm char}}{M}\right)^{1.6}\right]}\mbox{\ ,}
\end{equation}
where $M_{\rm char}=30\msun$ is the characteristic mass. Above $M_{\rm char}$, it follows a Salpeter-like IMF, but is exponentially cut off below that mass (e.g. \citealp{Chabrier2003}; \citealp{Wise2012}). The star formation efficiency for Pop~III stars in minihaloes is still debated (for a review, see \citealp{Bromm2013}). The minimum resolved mass of minihaloes in the simulations is $M_{\rm vir}\approx3\times10^6\msun$, which is rather massive compared to the typical minihalo mass of $M_{\rm vir}\sim 10^5-10^6\msun$. Extrapolating the star formation efficiency for Pop~III stars, $f_{\ast,\rm Pop~III}= M_{\ast}/M_{\rm b}\approx 2\times10^{-3}$ for typical minihaloes, we assign a total stellar mass of $M_{\ast}\approx1200\msun$ to the $M_{\rm vir}\approx3\times10^6\msun$ halo. Here, the baryon mass is $M_{\rm b}=M_{\rm vir}\times f_{\rm b}$, with a global baryon fraction of $f_{\rm b}=0.168$.
\par
As noted earlier, the resolution is not sufficient to follow the evolution of collapsing gas up to the highest densities, and hence 
we employ a sink technique (\citealp{Bromm2002}), where stellar clusters are represented by collisionless sink particles. Once a gas particle exceeds a pre-determined threshold density, $n_{\rm th}$, the highest-density SPH particle is converted into a sink particle, subsequently accreting neighboring gas particles, until the total mass in Pop~III stars reaches $M_{\ast, \rm Pop~III}=1200\msun$. For the threshold density, we here adopt $n_{\rm th}=220 \cmci$, where the corresponding free-fall time of the gas is $t_{\rm ff}\approx$3 Myr. We allow sink particles to merge once their distance 
falls below the SPH kernel size, $h_{\rm smooth}\approx 9 \unit{pc}$, comparable to the baryonic resolution scale of $l_{\rm res} \equiv [( 3X M_{\rm res})/(4 \pi n_{\rm th}
  m_{\rm H})]^{1/3} \approx 9 \pc$, where $X=0.76$ is the hydrogen mass fraction. Here, the baryonic mass resolution is 
  $M_{\rm res} \equiv N_{\rm ngb} m_{\rm SPH}\approx 2.4\times10^4\msun$, where $N_{\rm ngb} = 48$ is the number of particles in 
  the SPH smoothing kernel \citep{Bate1997}. The new positions and velocities of the merged sinks are assigned as mass-weighted mean values.
\par

\subsubsection{Population~II}
Second-generation Pop~II stars are formed out of metal-enriched gas, polluted by heavy elements ejected by preceding Pop~III SNe. We employ a star formation recipe similar to that for Pop~III, but adding an additional metallicity criterion: if the metallicity of a gas 
particle, eligible for star formation, exceeds the critical metallicity, $Z_{\rm crit}=10^{-5}\zsun$, for the transition from 
Pop~III to Pop~II, we assume that a newly formed sink particle represents a Pop~II star cluster (see \citealp{Bromm2013}). The star formation efficiency 
of Pop~II stars out of the metal-enriched clouds is also uncertain. For guidance, we turn to the study on Pop~II star formation in metal-enriched atomic cooling haloes by \citet{Safranek2016}, predicting that a pre-stellar clump in the centre of the forming galaxy is converted into a final stellar mass of 
$3000\msun$. Assuming that their results can be considered as representative for 
how stars form in all clumps, we form $M_{\ast, \rm Pop~II}=3000\msun$ out of a single metal-enriched molecular cloud. Hence, once a gas particle satisfies both criteria, exceeding the threshold density, $n_{\rm th}=220 \cmci$, and the critical metallicity, $Z_{\rm crit}=10^{-5}\zsun$, we immediately create an effective sink particle with a mass of $M_{\ast, \rm Pop~II}=3000\msun$ by accreting 
surrounding gas particles. For the Pop~II IMF, we employ a standard Salpeter form, $dN/d\log m\approx m^{-\alpha}$, with a slope 
$\alpha=1.35$ over the mass range of $[0.1-100]\msun$. 
\par

\subsection{Feedback physics}
\subsubsection{Photoionization feedback}
Once stars are created, star particles begin to emit ionizing photons which are propagated by solving the radiative transfer (RT) equations in TRAPHIC (\citealp{Pawlik2008,Pawlik2011}), as implemented in GADGET. Photon packets from radiation sources are transferred along the unstructured, spatially adaptive grid set by SPH particles in a photon-conserving manner. As such, photon packets originating in star particles are transported onto $\bar{N}_{\rm ngb}=32$ neighboring SPH particles, residing in $N_{\rm EC}=8$ tessellating emission cones. If the cones contain no SPH particles, so-called virtual particles are inserted to which the photon packet is transferred. To increase the sampling of the volume, photon packets are 
emitted $N_{\rm em}=\Delta t_{\rm r}/ \Delta t_{\rm em}$ times by randomly rotating the orientation of the cones, where 
the emission time step is $\Delta t_{\rm em}=10^{-2}$ Myr and the radiative timestep $\Delta t_{r}=\unit{min}(10^{-1} \unit{Myr}, 
\Delta t_{\rm hydro})$. Here, $\Delta t_{\rm hydro}$ is the minimum GADGET particle timestep. One of the prominent features 
of TRAPHIC is that the computational cost is independent of the number of ionizing radiation sources by virtue of a photon packet merging technique. The RT computation is self-consistently coupled to the hydrodynamical evolution such that the photoionization, 
photoheating, and photodissociation rates are computed in the RT module and used as inputs to the non-equilibrium solver for the chemical and thermal evolution of the gas. We refer the reader to \citet{Pawlik2008,Pawlik2011} for further details.
\par
Pop~III stars from a single stellar cluster emit ionizing photons at a total rate of 
\begin{equation}
\dot{N}_{\rm ion, tot} (t)=\int_{m_{\rm low}}^{m_{\rm up}}{\dot{N}_{\rm ion}(m)\phi_{\rm Pop~III}(m) \eta(m)}dm,
\end{equation}
with the same choice of Pop~III IMF parameters as above. Here, the contribution function is $\eta(m)=1$, as long as the age 
of a Pop~III cluster is less than the lifetime of an individual Pop~III star, $t_{\rm age} < t_{\ast}(m)$, and $\eta(m)=0$ after a star leaves the main-sequence. For the ionizing photon rates of individual Pop~III stars and their lifetimes, we use polynomial fits (\citealp{Schaerer2003}), $\log_{10} \dot{N}_{\rm ion}=43.61+4.90 x-0.83 x^2$ and $\log_{10} t_{\ast}=9.785-3.759x+1.413x^2-0.186x^3$, respectively, where $x=\log_{10}(m/\msun)$ refers to the initial mass of a star. For instance, the integrated ionizing photon rate on the zero-age main-sequence is $\dot{N}_{\rm ion, tot} (t=0)=10^{47.9} \rm s^{-1} \msun^{-1}$.
\par 
For Pop~II clusters, we use the ionizing photon rates from \citet{Schaerer2003} for various metallicities. Specifically, we here adopt the rates derived for a Salpeter IMF at a metallicity of $Z=5\times10^{-3}\zsun$, in the mass range $[1, 150]\msun$. At the main-sequence stage, the rate is 
$\dot{N}_{\rm ion}=10^{47} \rm s^{-1} \msun^{-1}$, but as massive stars die, it declines as the cluster evolves (\citealp{Schaerer2003}). We restrict the emission from Pop~II stars to last for 20 Myr, corresponding to the typical lifetime of OB stars in the cluster, beyond which the number of ionizing photons is insufficient to affect the gas.
\par
Additionally, we track the Lyman-Werner (LW) radiation from Pop~III and Pop~II clusters that dissociates molecular hydrogen ($\rm H_2$) and deuterated hydrogen (HD) in the optically thin limit, but considering a self-shielding correction in gas with a high hydrogen column density (\citealp{Wolcott2011b}). For the LW photon rate emitted by Pop~III stars, we employ the following fit, $\log_{10} \dot{N}_{\rm LW}=44.03+4.59x-0.77x^2$ (\citealp{Schaerer2002}). Note that we only form stars within the most refined region, comparable to a linear size of $\approx$ 450 comoving kpc, and we ignore the propagation of photons outside this region. 
\par

\begin{figure}
  \includegraphics[width=87mm]{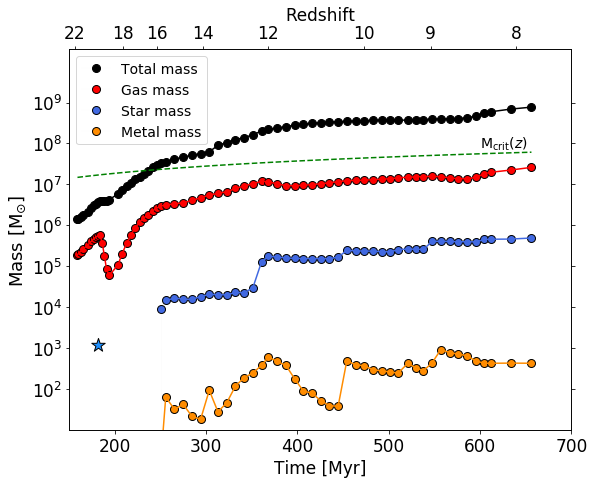}
   \caption{Assembly history of the emerging first galaxy, beginning with a minihalo at $z\gtrsim22$ with a mass of $M_{\rm vir}\approx10^6\msun$ and growing to $M_{\rm vir}\approx10^9\msun$ at $z\approx8$. 
   The critical halo mass, above which gas cooling is dominated by atomic hydrogen lines, is denoted by the dashed line as a function of cosmic time. The susceptibility of the minihalo to disruption is reflected in the gas evolution, such that more than 90\% of the gas within the host halo is evacuated by stellar feedback from metal-free stars at $z\approx19$ (marked by the star symbol). The second generation of stars emerges at $z\approx17$, out of the gas enriched by metals from Pop~III SNe. By the end of the simulation, the galaxy contains a stellar mass of $M_{\ast}\approx7\times10^5\msun$. Due to the complex interplay between metal ejection and re-accretion, the metal content of the galaxy oscillates and it reaches $m_{\rm metal}\approx320\msun$ at the end of the simulation.  \label{Fig:fg_evol}}
\end{figure}

\subsubsection{Supernova feedback}
It is well-known that SN feedback is one of the most efficient feedback mechanisms that alter the surrounding interstellar medium, destroy dense gas, and thus regulate star formation activity inside galaxies. The effectiveness of SN feedback depends on numerical resolution and on the details of its implementation, i.e., whether SN energy is inserted as thermal or kinetic energy. In this study, we use a thermal energy scheme: when massive stars in Pop~III and Pop~II clusters die as supernovae, a fraction of their rest-mass energy is converted into thermal energy, imparted to neighboring SPH particles. 

A well-known problem related to the thermal energy method is that neighboring gas particles that are heated by the SN explosion 
radiate their energy away too quickly, rendering SN feedback ineffective. This may in part be due to the absence of photoheating and thus a lack of photo-evacuation of gas prior to the SN explosion. The surrounding medium is then too dense, thus artificially boosting cooling losses. 

In our simulation, we avoid overcooling by explicitly accounting for photoheating prior to the SN explosion, and by decreasing the amount of gas that receives SN energy, as proposed by \citet{Vecchia2012}. Specifically, in order to attain a temperature jump of $\Delta T\gtrsim10^{7.5}$K, which is necessary to make SN feedback on the surrounding gas effective, we reduce the number of neighboring heated particles to a single particle, the nearest SPH particle surrounding a stellar cluster. In addition, we utilize a timestep-limiter in which the ratio of timesteps of neighboring SPH particles is designed to be smaller than a given factor, 4 in this study (\citealp{Saitoh2009}, \citealp{Durier2012}). This implementation is particularly important to avoid large integration errors that commonly occur in high resolution multiphase simulations with individual time-steps, as the heated hot gas ($T\gtrsim10^{7.5}$ K) is likely to be located near cold, dense gas ($T\lesssim10^4K$) at a SN explosion site. Also, we adopt a timestep-update (\citealp{Vecchia2012}), which allows neighboring particles that receive SN energy to immediately react to a sudden SN event by making them active particles at the time of energy injection.

The number of stars per unit stellar mass resulting in SNe is defined as
$n_{\rm SN} = \int_{m_0}^{m_1}{\phi(m)} dm$. Here, $m_0$ and $m_1$ are the minimum
and maximum initial mass of stars encountering a SN explosion, and $\phi(m)$ is a given IMF. For a Pop~III cluster, the number of CCSNe and pair-instability SNe (PISNe) per stellar mass are $n_{\rm CCSN, Pop~III}=1.3\times10^{-2}\msun^{-1}$ ($[m_0, m_1]=[11,40]\msun$) and $n_{\rm PISN, Pop~III}=1.8\times10^{-4}\msun^{-1}$ ($[m_0, m_1]=[140,150]\msun$), respectively. Applying the Salpeter IMF, the corresponding number of CCSNe in a Pop~II cluster is $n_{\rm CCSN, Pop~II}=7.42\times10^{-3}\msun^{-1}$ ($[m_0, m_1]=[8,100]\msun$). The total available energy from a single Pop~III SN event is $E_{\rm SN, Pop~III}= (\epsilon_{\rm CCSN, Pop~III} + \epsilon_{\rm PISN, Pop~III}) \times M_{\ast, \rm Pop~III}=1.78\times10^{52} \unit{erg}$. Here, the total available CCSNe and PISNe energies per unit stellar mass are $\epsilon_{\rm CCSN, Pop~III}=n_{\rm CCSN,Pop~III}\times 10^{51} \unit{erg}$ and $\epsilon_{\rm PISN, Pop~III}=n_{\rm PISN,Pop~III}\times10^{52} \unit{erg}$, respectively. For a Pop~II cluster, the total amount of SN energy is $E_{\rm SN, Pop~II}= \epsilon_{\rm CCSN, Pop~II}\times M_{\ast, \rm Pop~II}=2.2\times10^{52} \unit{erg}$, where $\epsilon_{\rm CCSN, Pop~II}=n_{\rm CCSN,Pop~II}\times 10^{51} \unit{erg}$. We instantaneously release the SN energy 3~Myr and 4~Myr after the formation of the clusters, employing typical lifetimes of massive Pop~III and Pop~II stars.

\par

When a SN explosion is triggered, ejected metals are evenly distributed onto neighboring particles, $N_{\rm ngb}=48$, resulting in the initial metallicity,
$Z_i = m_{\rm metal,i}/(m_{\rm SPH}+m_{\rm metal,i})$,
where $m_{\rm metal,i}=M_{\ast}y_{\rm eff}$. The IMF-averaged effective metal yields 
for the Pop~III and Pop~II CCSNe are $y_{\rm eff, Pop~III \hspace{0.05cm} CCSN}=0.05$ and $y_{\rm eff, Pop~II \hspace{0.05cm} CCSN}=0.005$, respectively, and we take $y_{\rm eff, Pop~III \hspace{0.05cm} PISN}=0.5$ for the energetic Pop~III PISNe (\citealp{Heger2002}). Due to the lack of mass flux between SPH particles, the implementation of metal transport in cosmological galaxy formation simulations is nontrivial. We employ a diffusion-based method for metal transfer developed by \citet{Greif2009}, where the mixing efficiency on unresolved scales is governed by the physical properties on the scale of the SPH smoothing kernel (\citealp{Klessen2003}). Here, the diffusion coefficient is defined as $D=2$ $\rho$ $\tilde{v}$ $\tilde{l}$, where the characteristic scale, $\tilde{l}$, is set by the smoothing length of the SPH kernel, $\tilde{l}=h$, and $\rho$
is the gas density. The velocity dispersion within the kernel,
$\tilde{v}$, is computed via
$\tilde{v}_i^2 = \frac{1}{N_{\rm ngb}} \sum_j |v_i-v_j|^2$, where $v_i$ and $v_j$ are the velocities of particles $i$ and $j$ within the kernel.

\begin{figure}
  \includegraphics[width=86mm]{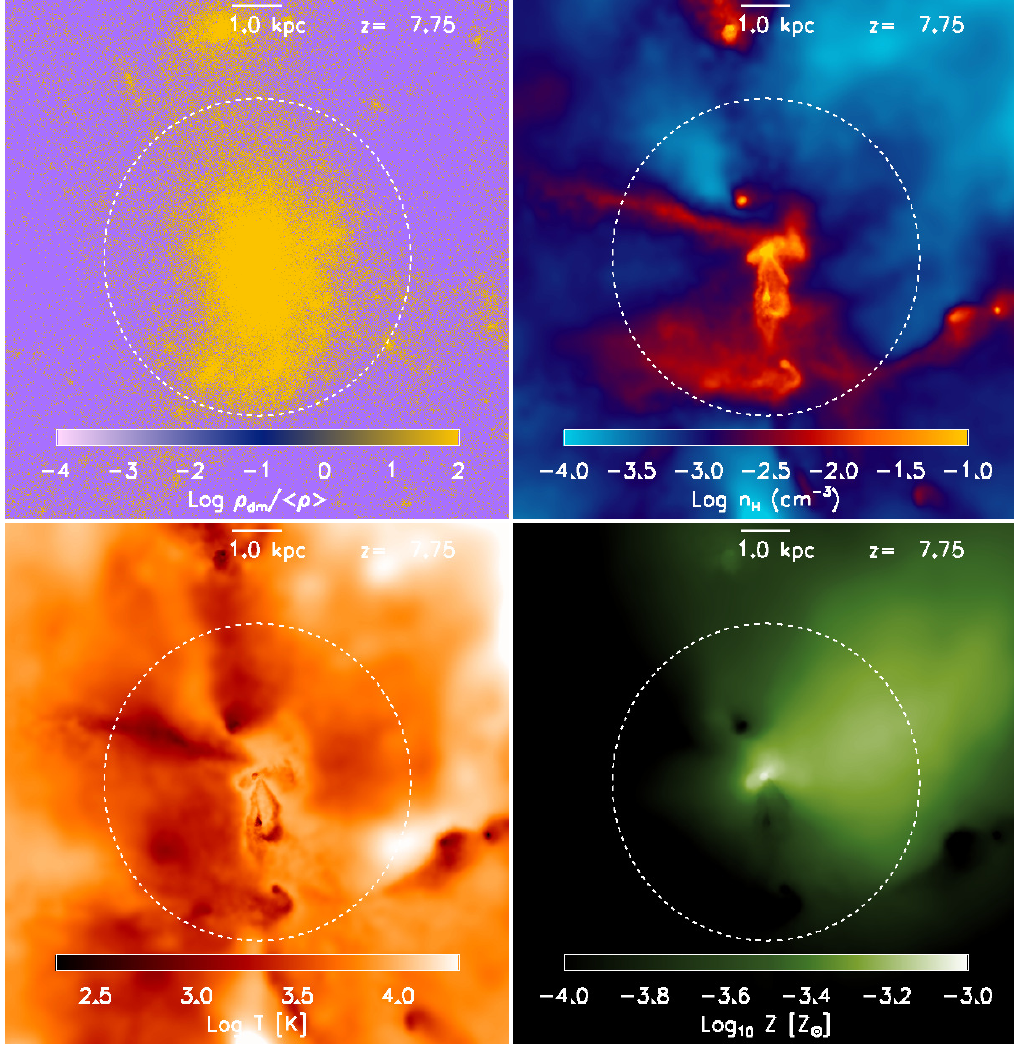}
   \caption{Morphology in the emerging first galaxy at the end of the simulation at $z\approx8$. {\it{Clockwise from upper left:}} Dark matter overdensity, hydrogen number density, gas metallicity, and gas temperature, projected along the line of sight within the central $\sim100$ comoving kpc. Dashed white circles at the center indicate the virial radius of the halo, which is $r_{\rm vir}\sim 3$ kpc at $z\approx8$. 
    \label{Fig:snap}}
\end{figure}

\section{Simulation results}

\subsection{Assembly history and galaxy morphology}

Fig.~1 shows the evolution of the simulated galaxy that begins assembling from a minihalo at $z\approx22$ and grows to 
$M_{\rm vir}\approx10^9\msun$ at $z\approx8$. An initial Pop~III cluster forms at $z\sim19$ in a minihalo with 
a mass of $M_{\rm vir}\approx2\times10^6\msun$. The photoionization heating, together with the following supernova 
feedback from the Pop~III cluster, leaves the minihalo sterile, losing $90\%$ of the total gas within its virial radius. As a consequence, further star formation is suppressed during the subsequent $\approx60$ Myr, such that the second generation of star formation is delayed until $z\approx15.6$. The minihalo continues to grow and exceeds the mass threshold for the onset of atomic hydrogen colling at $z\approx17$. While the halo becomes gradually less vulnerable to disruption with increasing virial mass, thus being able to retain its overall gas mass, stellar feedback still impacts the gas density near star-forming regions by episodically regulating the amount of the central gas. By the end of the simulation at $z\approx8$, the total stellar mass reaches $M_{\ast}\approx 7\times10^5\msun$. Chemical enrichment is achieved by metals from Pop~III and Pop~II SNe that are mixed into the ISM of the galaxy. Due to stellar feedback, a fraction of the enriched material is evacuated out of the host halo. On the other hand, ongoing star formation and accretion of pre-enriched gas replenishes its metal content. At $z\approx8$, the average metallicity of the galaxy is $Z\approx2\times10^{-3}\zsun$. 

\begin{figure}
  \includegraphics[width=86mm]{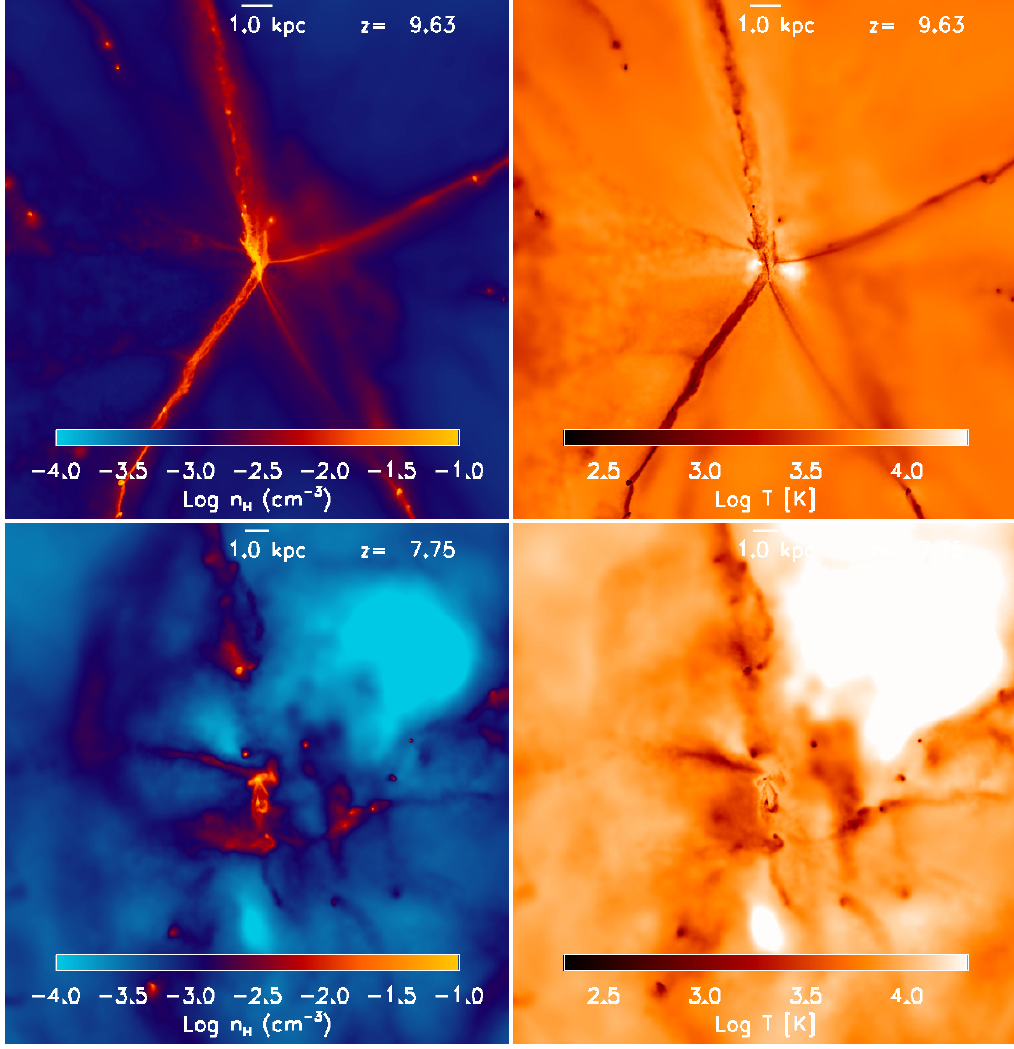}
   \caption{Effect of SN feedback on the gas in the first galaxy. Gas density ({\it left panels}) and temperature ({\it right panels}), averaged along the line of sight, in a cubical region of linear extent 250 comoving kpc, centered on the target halo in runs without ({\it top panels}) and with ({\it bottom panels}) SN feedback. Filamentary structures in the cosmic web are well preserved in the simulation without SN feedback, while the gas supply into the centre of the galaxy is disrupted otherwise, as clearly shown in the bottom panels.
    \label{Fig:compare}}
\end{figure}

\begin{figure}
  \includegraphics[width=87mm]{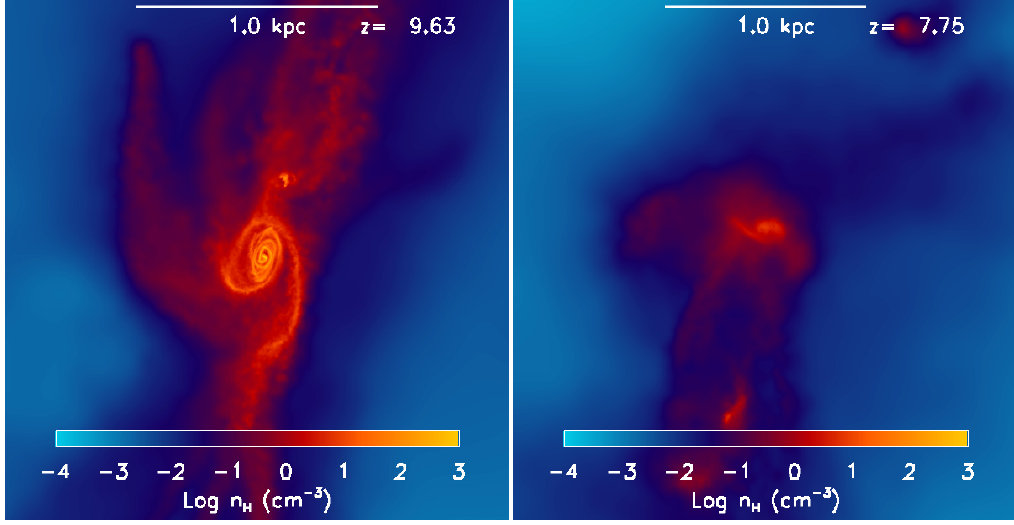}
   \caption{Survival of central disk structure. Face-on views of the gas density centered on the target galaxy in the simulations without ({\it left}) and with ({\it right}) SN feedback. In the left panel, a prominent disk structure is exhibited, while the disk is completely destroyed by the SN feedback ({\it right panel}). 
    \label{Fig:compare_disk}}
\end{figure}

The morphology of the simulated galaxy at $z\approx8$ is shown in Fig.~2, clearly displaying the inhomogeneous distribution of the baryonic component. This is caused by the stellar feedback preferentially driving the gas toward the direction perpendicular to cosmic filaments, lowering the gas density to $n_{\rm H}\lesssim10^{-3.5}$ $\rm cm^{-3}$. The metals expelled by subsequent SNe are therefore propagating along the channels created by the stellar feedback, as seen in the bottom-right panel of Fig.~2. The effect of stellar feedback, especially SN feedback, on the galaxy is evident in Fig.~3, where we compare two simulations, one without and one with SN feedback (top and bottom panels, respectively). The cosmic filaments that feed the gas onto the galaxy are disconnected by SNe feedback, thus impeding IGM gas from falling into the central star-forming region of the galaxy. To the contrary, without SN feedback the cosmic web filamentary structure is well preserved, constantly providing the galaxy with gas from the IGM. We point out that a gaseous disk is found within the inner region, $r\lesssim500$ pc, in the run without SN feedback, whereas such a structure is completely disrupted in the presence of SN feedback (see Fig.~4). The photoionization heating from stellar clusters is included in both runs, indicating that the existence of disk structure is not sensitive to photoionization heating, but strongly to SN feedback, in line with previous results \citep[e.g.][]{Pawlik2013}.

\begin{figure}
  \includegraphics[width=85mm]{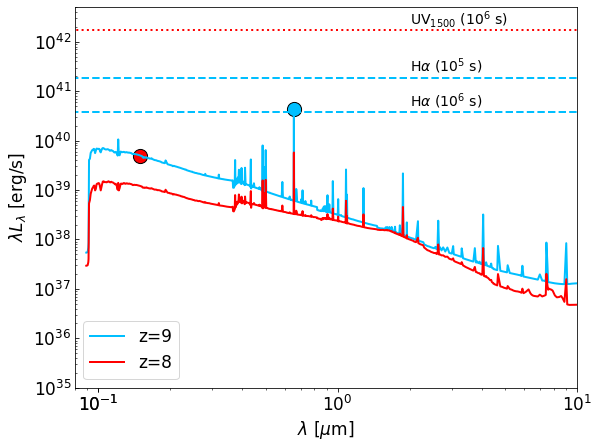}
   \caption{Post-processed integrated SEDs of the simulated galaxy at two characteristic evolutionary epochs: after experiencing a starburst with a star formation rate of $\approx0.1\msun \rm yr^{-1}$ at $z\sim9$ ({\it blue}), and at the end of the simulation at $z\sim8$ ({\it red}). We particularly mark the luminosities of UV continuum at $\lambda=$1500{\AA} and H$\alpha$ at $z=9$ as red and blue circles. The resulting spectra show that the simulated galaxy ($M_{\rm vir}\approx10^{9}\msun$ at $z=9$) may be observable using the H$\alpha$ recombination line with a deep exposure of $10^6$~s, while other emission lines, such as Ly$\alpha$ and He~II 1640, and the UV continuum at $\lambda=$1500{\AA} are below the detection limit of the {\sc JWST}, unless boosted by gravitational lensing.
    \label{Fig:Spec}}
\end{figure}

Here, we should point out the major caveat that our analysis is based on a single galaxy, thus not allowing us to consider halo-to-halo variance. In particular, low-mass galaxies are highly susceptible to differences in the stellar feedback they experience, resulting in different star formation histories and stellar masses. To assess the robustness of this work, we compare with other studies that have investigated the properties of high$-z$ galaxies over a similar mass range explored here.

Such a large comparison sample is provided by the {\it Renaissance} simulation (\citealp{Xu2016}), a zoom-in cosmological radiation hydrodynamics simulation with a DM mass resolution of $m_{\rm DM}\sim2.9\times10^4\msun$, providing statistical properties of galaxies in the mass range $M_{\rm vir}=10^7-10^{9.5}\msun$ at $z\gtrsim8$. Their analysis shows that for $M_{\rm vir}=10^{8.75}\msun$, similar to the halo mass in our work, the median value of stellar mass is $M_{\ast}\approx10^{6.24}\msun$, with a large scatter over the range $M_{\ast}=10^5-10^7 \msun$. Another simulation suite, performed by \citet{Ma2018} with comparably high resolution, derives a best-fit stellar to halo mass relation, resulting in a stellar mass of $M_{\ast}\approx 10^{5.2}\msun$, with a 1$\sigma$ dispersion of $\log M_{\ast}\simeq 0.39$, for a halo mass comparable to the system studied here. Interestingly, our $M_{\ast}-M_{\rm halo}$ value is close to the lower limit suggested by \citet{Xu2016}, while it is consistent with the upper limit of \citet{Ma2018}. Such differences between simulations can be attributed to the intrinsic halo-to-halo scatter, as well as to resolution and numerical details implemented in each simulation.

\subsection{Spectral energy distribution}
We post-process the simulated galaxy using a three-dimensional dust radiation transfer code, {\sc SUNRISE}, which uses a polychromatic algorithm in order to produce the spectral energy distribution (SED) of a galaxy. For detailed descriptions, we refer the reader to \citet{Jonsson2006} and \citet{Jonsson2010}. Briefly summarizing the key ingredients here, the code considers direct stellar light, nebular emission, as well as dust emission and absorption. Each star particle represents a stellar cluster that has its own age, metallicity, and mass, all of which is used as an input to the STARBURST99 code (\citealp{Leitherer1999}), assuming a Kroupa IMF with a lower and upper mass limit of 0.1$\msun$ and 100$\msun$, respectively. Within {\sc SUNRISE}, the photo-ionization code MAPPINGS~III is implemented (\citealp{Dopita2005, Groves2008}), calculating the nebular emission from H~II and photo-dissociation regions (PDRs). The dust mass is derived from the amount of metals associated with each SPH particle by adopting a dust-to-metal ratio of D/M\,$=0.4$. The dust temperature, emission, and opacity are obtained in a self-consistent fashion in {\sc SUNRISE} by iteratively solving the thermal equilibrium condition for dust grains at every location, where dust cooling is balanced by stellar radiative heating.

\begin{figure}
  \includegraphics[width=85mm]{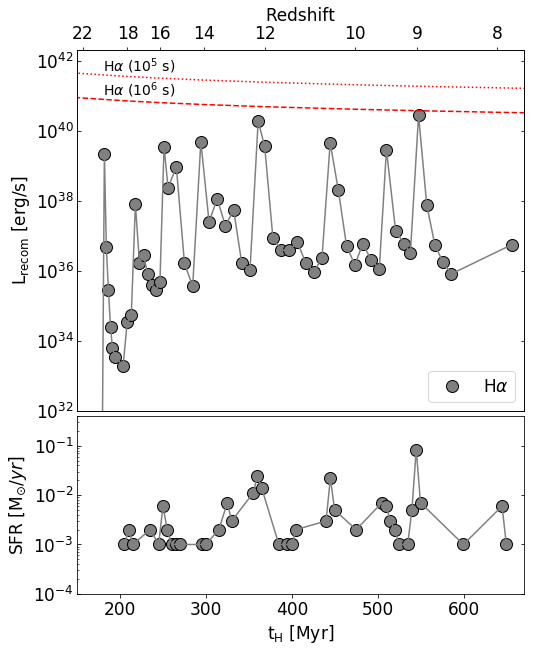}
   \caption{Evolution of recombination luminosity in the $\rm H\alpha$ line (top) and star formation rate of the galaxy (bottom). The luminosity is computed using Equ.~(3), with gas properties evaluated for individual SPH particles. Due to the bursty star formation in the simulated galaxy, recombination luminosities also episodically change, increasing as stars form and decreasing when the gas cools. We compare the derived recombination luminosity in $\rm H\alpha$ with the minimum value required to be detected with the JWST. When the galaxy experiences an active starburst ($\dot{M}_{\ast}\approx0.1\msun \rm yr^{-1}$), as shown in the bottom panel at $z\approx9$, the strength of H$\alpha$ nebular emission can be comparable to the detection limit with a deep exposure time of $10^6$ s.}
\end{figure}

Fig.~5 shows the resulting integrated spectra of the simulated galaxy for two characteristic times, when a starburst has just occurred at $z\approx9$ ({\it blue line}) and at $z\approx8$ ({\it red line}), which corresponds to the end of the simulation. 
At $z\approx9$, the star formation rate of the galaxy is $\dot{M}_{\ast}\approx0.1\msun \rm yr^{-1}$, and Pop~II stars with a total stellar mass of $M_{\rm \ast}\approx5.8\times10^5\msun$ reside in the halo, with $\sim$30\% newly formed during the recent starburst. We note that there are no young Pop~III stars present at $z\approx9$, as the star formation mode in the dense star-forming regions had already switched from Pop~III to Pop~II stars at $z\sim17$. For guidance, we provide select {\sc JWST} detection limits for the UV continuum at 1500{\AA} for a deep exposure of $10^6$~s, and for H$\alpha$ line emission for two exposure times, $10^5$ and $10^6$~s. 

During the starburst at $z\approx9$, the recombination luminosity in $\rm H\alpha$ can be comparable to the {\sc JWST} detection limit, assuming an exposure time of $10^6$~s and a signal-to-noise of $\rm S/N=10$. On the other hand, other line luminosities, including Ly$\alpha$ and He~II 1640\AA, and also the luminosity of the UV continuum at a wavelength of 1500{\AA} are all below the detection limit at both redshifts. It has been suggested that the nebular emission in the He~II 1640{\AA} line could be a sensitive indicator for metal-free Pop~III stars (e.g., \citealp{Bromm2001a, Schaerer2002, Schaerer2003}). The simulated galaxy, however, is already populated with metal-enriched stars, making the He~II 1640{\AA} line less prominent. The amount of dust within the virial volume of the simulated galaxy at $z\approx9$ is $m_{\rm dust}\approx120\msun$, given the dust-to-gas ratio of $\rm D/M=0.4$, where we adopt the value of the local Universe. We find that the reprocessed spectra are barely attenuated by the dust in the ISM. This is consistent with the results of recent observations, suggesting that high redshift galaxies with $M_{\rm vir}\lesssim10^{10}\msun$ are likely to show no or a little dust attenuation (e.g. \citealp{Finkelstein2012, Bouwens2014, Jaacks2017}).

Whether the simulated galaxy could be detected via nebular emission lines depends on its ISM gas properties. If the ionizing photons emitted from stars are trapped by the neutral ISM, they can be re-processed into recombination lines. On the contrary, ionizing photons might freely escape into the IGM, thus not contributing to the line emission. This dichotomy can be expressed with a covering factor, $f_{\rm cov}$, the fraction of the sky covered by neutral hydrogen gas (e.g. \citealp{Zackrisson2011}). Thus, $f_{\rm cov}=0$ indicates a situation where the gas clouds near newly formed stars are cleared up or fully ionized, giving rise to little nebular emission, while this emission would be maximized for $f_{\rm cov}=1$, describing an environment where stars are fully enclosed by neutral, dense gas clouds. 
\begin{figure}
  \includegraphics[width=87mm]{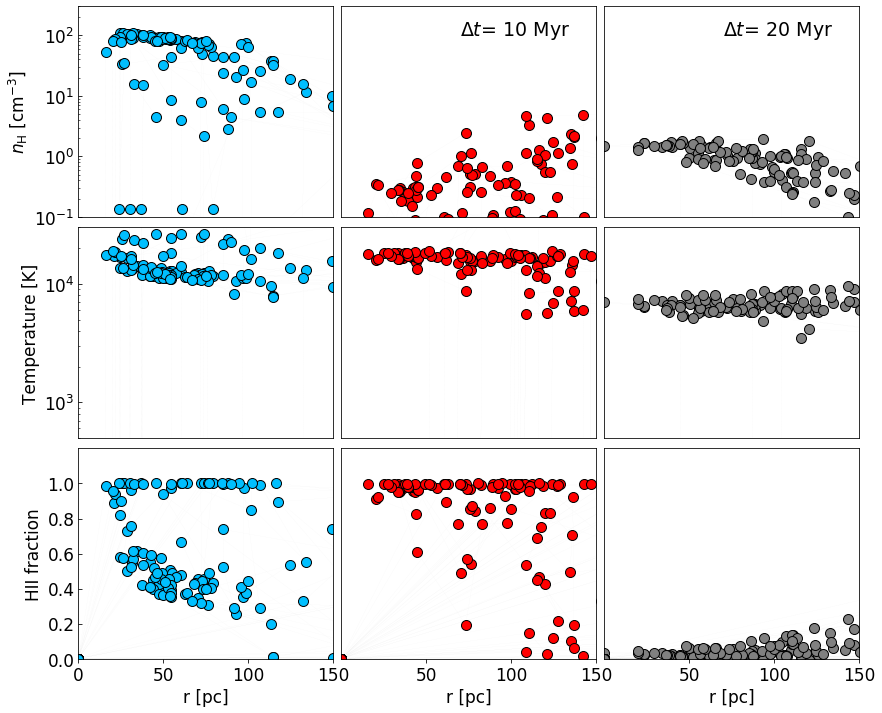}
   \caption{Evolution of circum-starburst gas properties. Shown are the hydrogen number density ({\it top}), gas temperature ({\it middle}), and ionized fraction ({\it bottom}), as a function of distance from the starburst at $z\approx9$. 
   Immediately after the starburst, displayed in the left panels, the density remains as high as $n_{\rm H}\approx 100 \mbox{\ cm}^{-3}$, but at the same time the gas is significantly heated to $T \sim10^4$~K. About 10~Myr after the burst, the gas has hydrodynamically responded to the photoionization heating, with central densities dropping to below $n_{\rm H}\sim 1 \mbox{\ cm}^{-3}$ within 100~pc.}
\end{figure}
\begin{figure}
  \includegraphics[width=85mm]{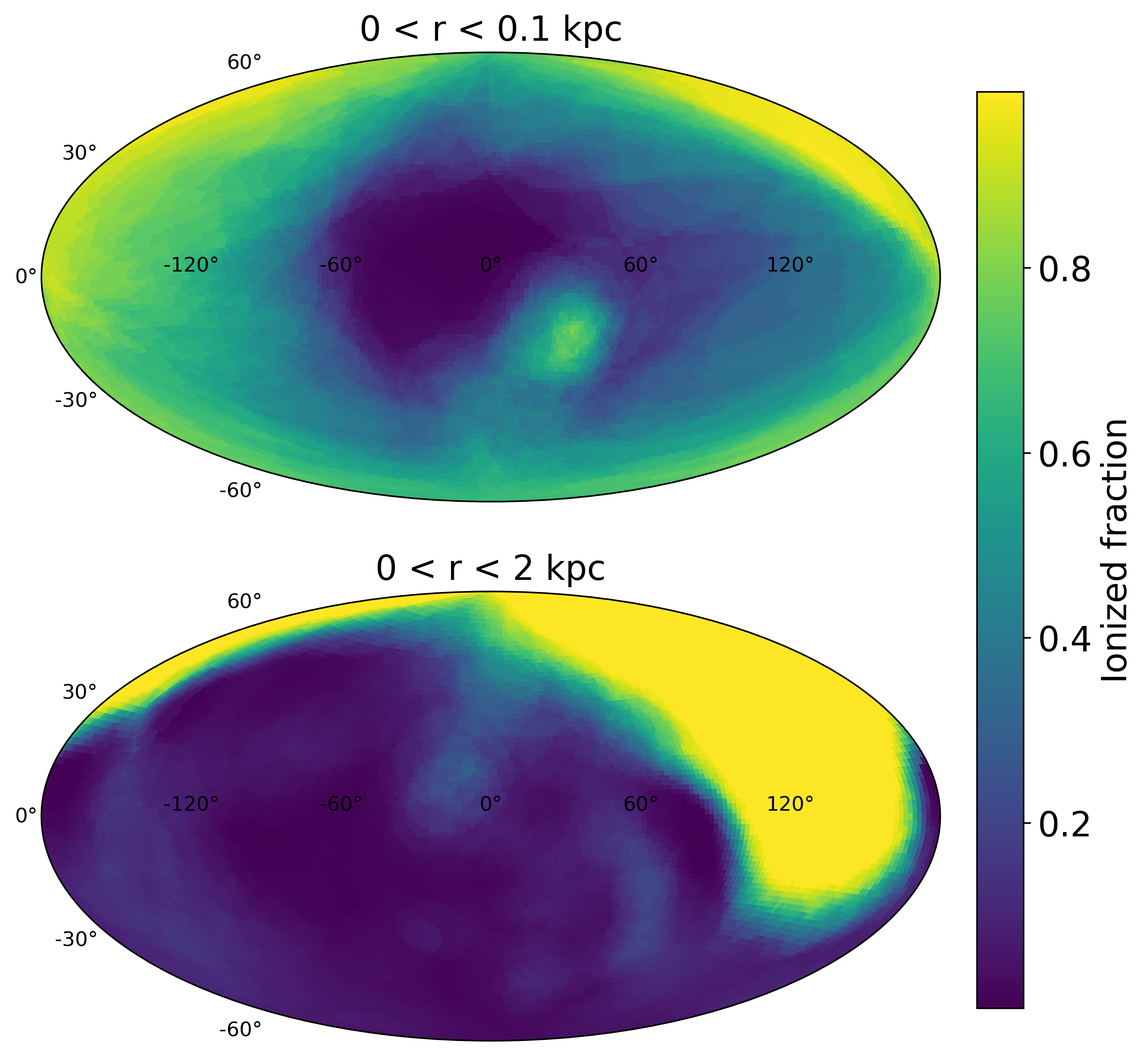}
   \caption{All-sky projection map of the ionized fraction within $r\lesssim100$~pc ({\it top}) and $r\lesssim 2$~kpc ({\it bottom}), immediately after the starburst at $z\approx9$. The pronounced anisotropy is clearly visible, imprinted by the inhomogeneous escape of ionizing photons into the surrounding cosmic web.}
\end{figure}
The gas conditions continue to change with ongoing star formation, resulting in a highly variable nebular emission. Given the inclusion of photoionization heating followed by SN feedback in our simulations, we self-consistently trace the evolving gas properties, and compute the resulting recombination lines directly from gas particles. In Fig.~6, we show the evolution of the star formation rate ({\it bottom}) and H$\alpha$ recombination line luminosity ({\it top}), emitted from photoionized gas within the galaxy, which is computed using the following equation, 
\begin{equation}
        L_{\rm H\alpha} = \sum_i {j_{\rm H\alpha} \frac{m_i}{\rho_i} \left[ \frac{\rho_i}{\mu_i m_{\rm H}}\right]^2 f_e f_{\rm HII}}\mbox{\ ,}
\end{equation}

where $j_{\rm H\alpha}$ is the temperature-dependent emission coefficient for the $\rm H{\alpha}$ line (\citealp{Osterbrock2006}). The individual gas mass, density, electron fraction, and ionized hydrogen fraction ($m_i$, $\rho_i$, $f_e$,  $f_{\rm H~II}$), are taken from each SPH particle within the virial radius of the galaxy. 

In general, the evolution of the recombination line emission tends to be consistent with the cycle of star formation activity in the galaxy, as the photoionized regions are generated around newly formed stars. Consequently, recombination luminosities change episodically such that they increase with rising star formation rates. The minimum luminosities to be observable with the {\sc JWST} for exposure times of $10^5$ and $10^6$~s decline with decreasing redshift. Throughout cosmic time, the simulated galaxy will be only detectable when it experiences the active starburst at $z\approx9$. The time interval for maximum nebular emission remains short ($\lesssim$ 2 Myr), confined to the earliest stage of starbursts, since the gas in the vicinity of the starburst is pushed outward as the H\,II region grows, lowering the central gas density. About $2-3$~Myr after the burst, the high-density gas is likely to be completely cleared up by the following SN feedback, thus significantly decreasing the strength of the nebular emission by $3-4$ orders of magnitude and allowing the leakage of ionizing photons into the IGM.

We find that the bursty nature of star formation likely plays an important role in the observability of high-$z$ galaxies, since an intense starburst is required for such galaxies to be detected in nebular emission. Such episodic star formation due to stellar feedback in low-mass galaxies has also been found in other simulations, capable of resolving individual star-forming regions (e.g. \citealp{Hopkins2013, Agertz2015, Dominquez2015, Sparre2017, Ma2018}). As a specific example, \citet{Hopkins2013} have suggested that the star formation efficiency, in a spatially- and time-resolved manner, can be self-regulated by stellar feedback, instead of imposing a fixed efficiency in subgrid models.  Similarly, \citet{FaucherGigu2018} has proposed a model predicting that the star formation should be bursty rather than constant in time, for low-mass galaxies and at high redshifts ($z\gtrsim1$). The reason is that the equilibrium between stellar feedback and gravity, resulting in the observed Kennicutt-Schmidt relation (\citealp{Kennicutt1998}), can break down in such cases.

Next to the H$\alpha$ line, other nebular emission lines can be feasible targets to probe high-redshift galaxies. In particular, far-IR fine structure lines, such as [O~III] 88 $\mu$m, are a reliable SFR tracer in low-metallicity dwarf galaxies (e.g. \citealp{DeLooze2014}). Extending this technique to high-$z$, recent observations with the Atacama Large Millimeter/submillimeter Array (ALMA) report the discovery of distant galaxies at $z=7-9$, using the [O~III] 88 $\mu$m line (e.g. \citealp{Inoue2016, Carniani2017, Laporte2017, Hashimoto2018, Tamura2018}). Such observations, however, are biased toward detecting massive galaxies ($M_{\ast}\gtrsim10^9\msun$ at $z>7$). An alternative rest-frame optical line, accessible to the JWST, is [O~III] 5007 {\AA} \citep{Moriwaki2018}, with an order of magnitude larger luminosity compared to the [O~III] 88 $\mu$m line. Specifically, the detectable lower limit for the [O~III] 5007 {\AA} line luminosity from a source at $z=8.3$ is expected to be $L_{\rm O~III, 5007}\sim4\times10^{41} \mbox{\ erg\,} \rm s^{-1}$, using the NIRCam narrow-band F466N filter with an exposure of $10^4$~s at a signal-to-noise of $\rm S/N=5$ (e.g. \citealp{Gardner2006, Greene2017, Moriwaki2018}). Rescaling this flux threshold to $10^6$~s, assuming $f_{\rm lim}\propto t_{\rm exp}^{-1/2}$, the [O~III] 5007 {\AA} line limit could be lowered by an order of magnitude. However, the estimated [O~III] 5007 {\AA} luminosity for our simulated galaxy is $L_{\rm O~III, 5007}\approx8\times10^{39}\mbox{\ erg\, s}^{-1}$, which is below the detection limit by a factor of $\gtrsim5$.

We should note that our derived [O~III] line luminosity is computed by assuming solar abundance ratios within {\sc Sunrise}. At early cosmic times, however, the ISM metal enrichment could be enhanced in $\alpha$-elements, ejected from CCSNe, or even PISNe, associated with early generation stars. Such enhancement could boost the visibility of high-$z$ [O~III] emitters (e.g. \citealp{Karlsson2013, Steidel2016, Katz2018}). We may thus underestimate the [O~III] 5007 {\AA} line luminosity in the present work, also due to the unusually low overall metallicity in our target galaxy. Tracing individual metal species in the ISM of high-$z$ galaxies in order to better estimate the strength of metal lines will be worth future investigation (e.g. \citealp{Pallottini2015}).

\subsection{Gaseous conditions}
As previously discussed, the ISM conditions around star-forming regions are one of the key factors that determine the importance of nebular emission. In Fig.~7, we exhibit the evolution of relevant gas properties as a function of central distance, specifically the hydrogen number density, gas temperature, and ionized fraction. We contrast the situation at the time of the starburst at $z\approx9$ ({\it left panels}) with the gas evolution $\sim10$~Myr and 20~Myr after the burst. Immediately after the starburst, most of the surrounding gas still remains at high density ($n_{\rm H}\sim100$ $\rm cm^{-3}$), while being substantially heated to $T\gtrsim10^4$~K, because it takes time for the gas to hydrodynamically response to the photoionization heating. After $\sim$10 Myr, shown in the middle panels of Fig.~7, central densities at $r\lesssim 50$~pc have fallen to $n_{\rm H}\lesssim1$ $\rm cm^{-3}$, and the gas is fully ionized. It takes another $\sim$ 10 Myr for the heated and ionized gas to cool and recombine again, resulting in ionized fractions of less than 10$\%$ within 100~pc. We note that gas properties significantly vary at a given distance, in particular the H\,II fraction. The inhomogeneous gas distribution is clearly evident in Fig.~8, where we show an all-sky projection map of the ionized fraction for the gas within 100~pc ({\it top}) and 2~kpc ({\it bottom}) at $z\approx9$. This indicates that photons  preferentially propagate along the direction perpendicular to cosmic filaments (see also Fig.~3).

In order to scrutinize further the behaviour of the neutral-gas covering factor near the starburst, in Fig.~9 we present the fraction of gas exceeding a given density threshold, as a function of distance from the centre of the galaxy about $\sim 5$~Myr before ({\it left panels}) and after ({\it right panels}) the starburst at $z\approx9$. We also show the corresponding covering factors of neutral hydrogen (H\,I) in the bottom panels of Fig.~9. The H\,I covering factor is defined as the fraction of the sky occupied by gas with a neutral fraction larger than a given threshold value. As shown in the top panels, $\sim 5$~Myr before and after the starburst, the gas density profiles show a similar trend; within $\sim 50$~pc, all the gas is dense with $n_{\rm H}\gtrsim 10$ $\rm cm^{-3}$, and the fraction of high density gas decreases with increasing distance, leading to the absence of such high density gas at $r\gtrsim 250$~pc. The covering factor, on the other hand, is strikingly different; $\sim5$~Myr prior to the starburst, all the gas at $\lesssim 200$~pc is fully neutral, such that $\sim$99$\%$ of the gas is in the neutral phase with a covering factor near unity. After the starburst, the gas at the centre of the galaxy ($r\lesssim200$ pc) becomes almost fully ionized, leading to a near-zero covering factor. We point out that the overall density of the gas near the starburst remains high, while the gas is significantly ionized. This is because the ionized gas has not yet responded hydrodynamically to the photoionization heating, and therefore the ionizing photons are initially trapped within the dense gas clouds surrounding the newly formed stars, resulting in the maximal nebular emission. 

\begin{figure}
  \includegraphics[width=85mm]{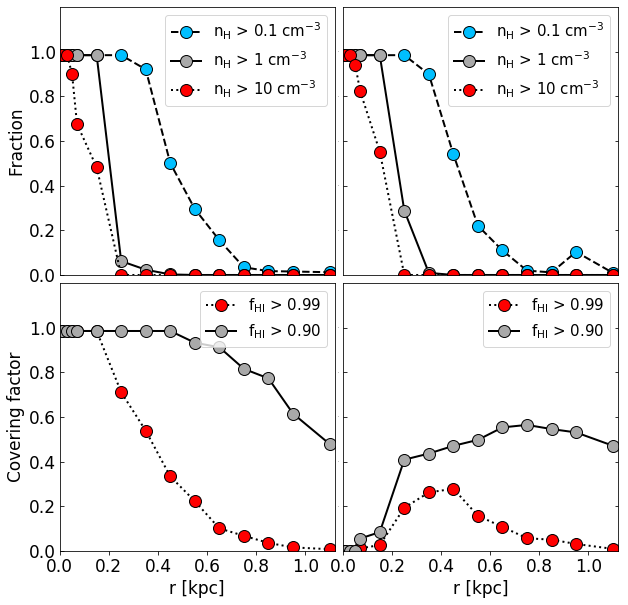}
   \caption{Evolution of neutral gas near the starburst. We show the fraction of gas exceeding select density thresholds ({\it top panels}), and the covering factor of neutral gas ({\it bottom panels}) $\sim 5$~Myr prior ({\it left}) and after ({\it right}) the starburst at $z\approx9$. 
   The covering factor is evaluated for neutral hydrogen fractions larger than $f_{\rm HI}=0.99$ and 0.9. Before the starburst, the gas is fully neutral, reflected in a 
   covering factor near unity. This early presence of high-density, substantially neutral gas results in the maximum contribution to the nebular emission.}
\end{figure}

\begin{figure}
  \includegraphics[width=85mm]{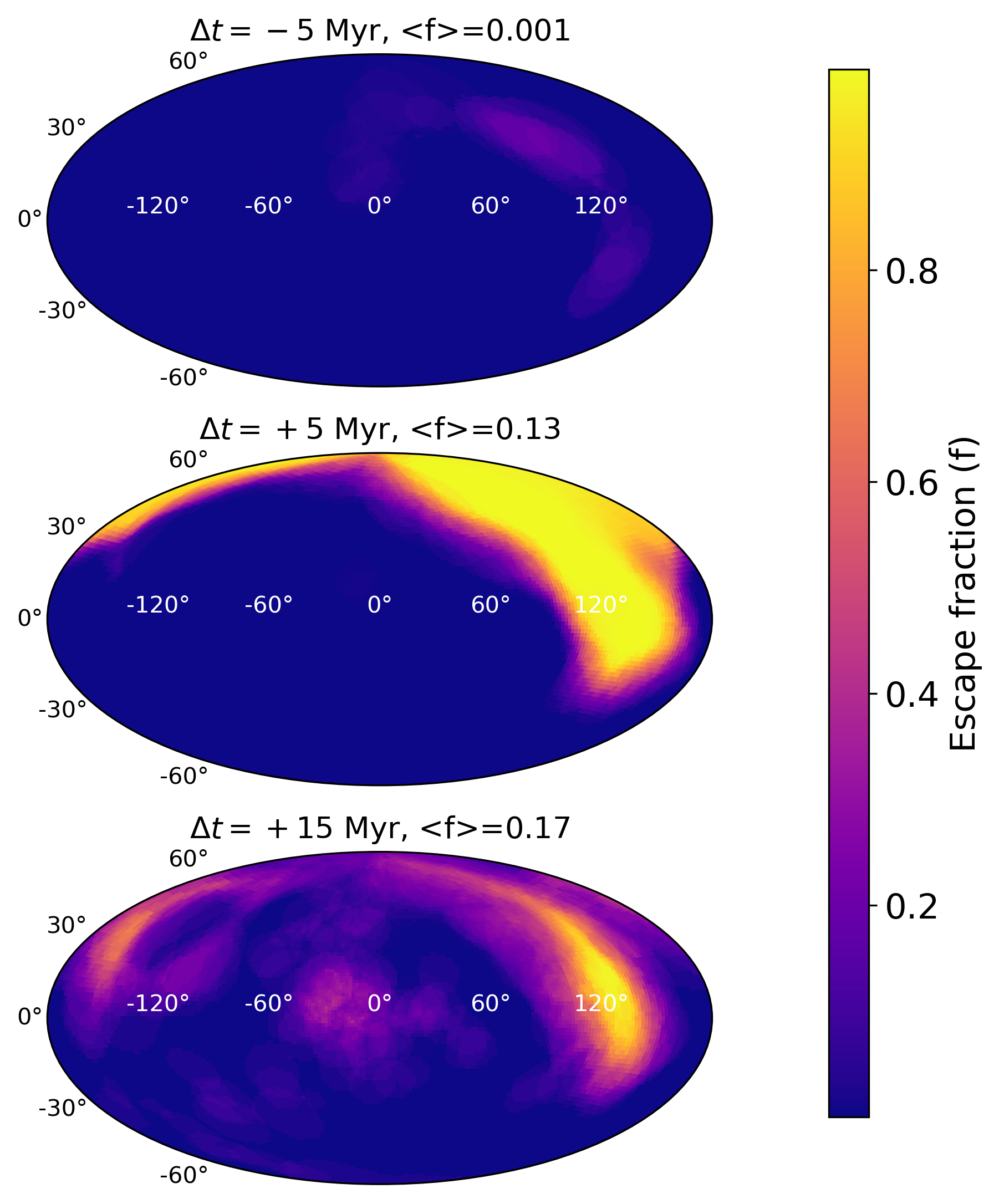}
   \caption{All-sky projection map of the escape fraction 5~Myr before the starburst at $z\approx 9$ ({\it top}), 5~Myr afterwards ({\it middle}), and 10~Myr after it ({\it bottom}). Prior to the starburst, most gas remains neutral (see also the left panels of Fig. 9), giving rise to near-zero escape fractions. About 5~Myr after the starburst, the average escape fraction rises to 13\%, corresponding to the gas conditions shown in the right panels of Fig.~9. We note that while photons first escape through the openings in the northeast direction, such low-density channels for ionizing photons to penetrate are more prevalent at later epochs, as displayed in the bottom panel. About 15~Myr after the starburst, the average escape fraction is $\sim$17\%.}
\end{figure}

During the starburst, the galaxy continues to form stars, such that 5~Myr later the total mass of the newly formed stars is $M_{\ast}\approx2\times10^5\msun$. Given that SN feedback ensues $\sim 2-3$~Myr after the formation of a stellar cluster, the sole action of photoionization heating is insufficient to decrease the central density and create channels for ionizing photons to escape into the IGM. This result is in agreement with previous findings (e.g. \citealp{Stinson2007}), pointing out the importance of SN feedback in destroying gas clumps in the vicinity of stars, thus enabling the leakage of ionizing photons from low-mass galaxies. In our simulation, the central gas density eventually drops to $n_{\rm H}\lesssim5 $ $\rm cm^{-3}$ over the next 10~Myr owing to the combination of photoionization heating and SN feedback, rendering the nebular emission inefficient. The time period, during which nebular emission is maximized, is thus less than $\sim 3$~Myr. 

In Fig.~10, we present the all-sky projection map of the escape fraction of ionizing photons before and after the starburst at $z\approx9$. The escape fraction is computed by dividing a sphere at the virial radius into 768 equal-area pixels using {\sc HEALPIX} (\citealp{Gorski2005}). Along each line of sight, we define a transmittance via $T=\exp(-\sigma_{\rm H~I} N_{\rm HI})$, where $\sigma_{\rm H~I}=1.78\times10^{-18} \rm cm^2$ is the H~I cross section at 21.6 eV, and $N_{\rm HI}$ the H~I column density along the line of sight from the radiation source to the virial radius. For simplicity, we assume that all radiation sources are located at the halo centre. Given the uniformity of pixel size, the average escape fraction is simply defined as the average transmittance over all pixels. As shown in the top panel of Fig.~10, before the starburst the average escape fraction is close to zero, as most of the gas in the galaxy is neutral (see also the left panels of Fig.~9). On the other hand, $\sim$ 5~Myr after the starburst, ionizing photons can penetrate the ISM and escape into the IGM along low-density, ionized channels, created by photoionization heating and SN feedback, resulting in an average escape fraction of 13\%. Therefore, the nebular emission will be reduced by a factor of $(1-f_{\rm esc})$, but a large number of ionizing photons are still trapped in the ISM, powering the line luminosity. More channels are available in different directions at later epochs, and the average escape fraction increases to $\sim$17\% at 15~Myr after the starburst, marking its maximum value.

\subsection{Observability}

One of the main targets of the {\sc JWST} is observing the first light from high-$z$ galaxies (\citealp{Gardner2006}). Onboard the {\sc JWST}, the Near Infrared Camera (NIRCam) is designed to observe the first light using deep field imaging with a field of view of $2.2'\times4.4'$ and an angular resolution of $\sim0.03''-0.06''$ in the range of observed wavelengths $\lambda_{\rm obs}=0.6-5\mu m$. However, non-ionizing radiation from the simulated galaxy, especially at 1500{\AA}, is unlikely to be detectable. Instead, the H$\alpha$ line could possibly be observed using the Mid Infrared Instrument (MIRI), which operates in the wavelength range of $\lambda_{\rm obs}=5-28.8 \mu m$, enabling imaging with a field of view and angular resolution of, respectively, $\sim2'\times2'$ and $\sim0.1''-0.6''$, as well as low and medium resolution spectroscopy. 

\begin{figure}
  \includegraphics[width=87mm]{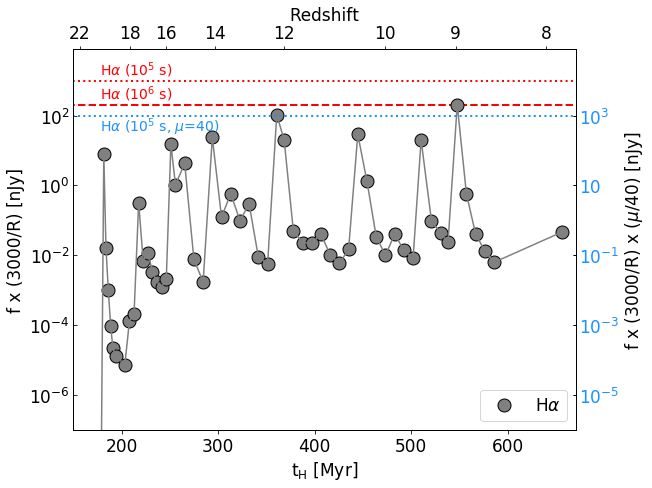}
   \caption{Evolution of the observed flux of the H$\alpha$ recombination line from the simulated galaxy. We compare to the detection limit of MIRI onboard the {\sc JWST} for a signal-to-noise ratio of $\mbox{S/N}=10$ and an exposure time of $10^5$ and $10^6$~s. The {\sc JWST} detection limit can be lowered by an order of magnitude, if the flux is boosted by gravitational lensing, assuming a mean magnification for a foreground galaxy cluster of $\mu\approx 40$. The evolution of the H$\alpha$ emission follows the cycle of star formation and feedback activity, such that the luminous phases only last for about $2-3$~Myr.}
\end{figure}

We find that the spatial size of the simulated galaxy is compact, such that $\sim88\%$ of stars are concentrated within $r\lesssim 500$~pc, corresponding to $r\approx 0.2 r_{\rm vir}$. The observed angular size of the region from which the nebular radiation is emitted can be estimated as $\Delta \theta=\Delta l / d_{\rm A}\approx 0.1''(\Delta l / 0.5 \hspace{0.01cm} {\rm kpc})[(1+z)/10]$, where the angular diameter distance is $d_{\rm A} = (1+z)^{-2} d_{\rm L}$, and the luminosity distance at redshift $z$ is $d_{\rm L}\approx100[(1+z)/10]$~Gpc. Thus, the emission line flux is unlikely to be spatially resolved, given that the angular size of the galaxy, $\Delta \theta \approx0.1''$, is similar to the angular resolution of MIRI ($0.1''-0.6''$). The observed monochromatic flux in H$\alpha$ from a spatially unresolved galaxy with a spectral resolution of $R=\lambda/\Delta \lambda$ is provided by (e.g. \citealp{Oh1999, Johnson2009, Pawlik2011})
\begin{eqnarray}
f_{\rm H\alpha} (\lambda_{\rm obs})  &=& \frac{l_{\rm H\alpha} \lambda_{\rm H\alpha} (1+z) R}{4 \pi c d_{\rm L}^2 (z)} \\
 &\approx& 60 \hspace{0.05cm} {\rm nJy} \left(\frac{l_{\rm H\alpha}}{10^{40} {\rm erg \hspace{0.01cm} s^{-1}}}\right) \left( \frac{1+z}{10}\right)^{-1} \left( \frac{R}{3000} \right),
\end{eqnarray} 
where $l_{\rm H\alpha}$ is the H$\alpha$ recombination line luminosity, $\lambda_{\rm H\alpha}=6563$~\AA \hspace{0.5cm} the rest-frame wavelength, and $\lambda_{\rm obs}=(1+z)\lambda_{\rm H\alpha}$ the observed wavelength. As discussed previously, the UV continuum luminosity, at rest wavelength $\lambda_{\rm em}=1500$~\AA, from a galaxy with $M_{\rm vir}\approx10^9\msun$ at $z\gtrsim8$ is expected to be below the {\sc JWST} detection limit. On the other hand, following the starburst at $z\approx9$, the H$\alpha$ recombination line is bright enough to be detectable with MIRI, with a signal-to-noise ratio of $\rm S/N = 10$ and a deep exposure time of $10^6$~s, as shown in Fig.~11.

The detectability of small galaxies at high-$z$ may increase if their fluxes can be boosted by foreground, low-redshift, galaxy clusters (e.g. \citealp{Bradley2008, 
Zheng2009, Hall2012}). For instance, \citet{Zackrisson2012} have explored the effect of gravitational lensing on the observability of high-$z$ galaxies, mainly consisting of Pop~III stars, using the lensing field, J0717.5+3745 (J0717) at $z\approx0.5$, one of the best lensing clusters known. They have suggested that the lensed {\sc JWST} survey can push the detection limit of high-$z$ faint galaxies in terms of star formation efficiency by $\approx$0.4 dex (3~h exposures) and $\approx$1.0 dex (30~h exposures), compared to an unlensed survey with a 100~h exposure. Assuming a mean magnification of $\bar{\mu}\approx40$ for the J0717 field, the minimum flux for the detection of the simulated galaxy would be lowered by an order of magnitude, depicted as the blue dotted line in Fig.~11. We, however, caution that accurate modeling of the lensing field will be required in order to provide precise estimates for the effect of gravitational lensing.

As displayed in Fig.~11, the time period over which the nebular emission is prominent only extends for less than $2-5$~Myr due to the bursty star formation activity in the small galaxy at high-$z$. Considering the short duration of nebular emission in small galaxies, we compute the total number of galaxies, $N(>z)$, detectable in H$\alpha$ recombination radiation in a single MIRI field of view, as displayed in Fig.~12. The number of galaxies per unit solid angle at $\gtrsim z$ is as follows, 
\begin{equation}
\frac{dN}{d\Omega} (> z) = \int_z^\infty dz' \frac{dV}{dz'd\Omega} \frac{\Delta{t_{\rm neb}}}{t_{\rm H}(z')} \int_{\rm M_{\rm min}(z')}^{\rm M_{\rm max}} dM \frac{dn(M,z')}{dM}\mbox{\ ,}
\end{equation}
where $t_{\rm H} (z)$ is the cosmic time at redshift $z$, $dV=c H(z)^{-1} d_{\rm L}^2 (1+z)^{-2}$ the comoving volume, and $H(z)=H_{\rm 0}[\Omega_{\rm m} (1+z)^3
+\Omega_{\Lambda}]^{1/2}$ the Hubble parameter, where, $H_{\rm 0}=71 \mbox{\ km\,s}^{-1} \mbox{\,Mpc}^{-1}$. For the differential comoving halo number density, $dn(M,z)/dM$, we adopt the halo mass function provided by \citet{Warren2006}. The minimum halo mass, above which the galaxy is likely to be detected with nebular emission from a starburst, is chosen as $M_{\rm min}\approx10^{8.8}\msun$, the mass of the galaxy simulated in this work at $z\approx8$. 

\begin{figure}
  \includegraphics[width=85mm]{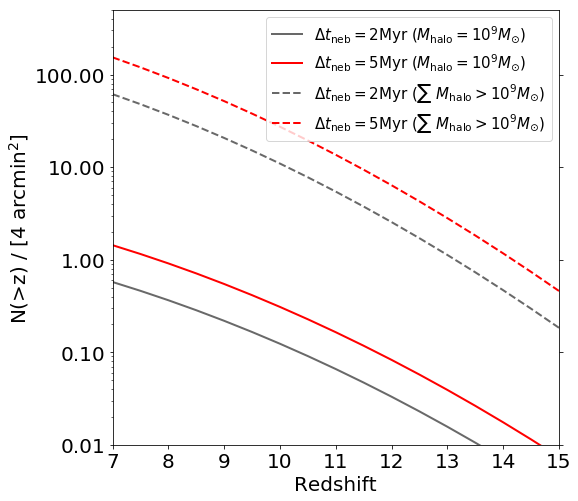}
   \caption{Number of galaxies detectable in the H$\alpha$ recombination line, using MIRI onboard the {\sc JWST} within a field of view of 2"x2". We consider two cases for the number of observable galaxies: one with a fixed mass of $M_{\rm vir}\approx10^{8.8}\msun$ ({\it solid lines}), corresponding to the mass of the simulated galaxy at the time of the starburst ($z\approx9$), and one for the total number of detectable galaxies above $M_{\rm vir}\gtrsim10^{8.8}\msun$ ({\it dashed lines}). Additionally, we consider two values for the duration of prominent nebular emission, $\Delta t_{\rm neb}=2$~Myr and $5$~Myr. We infer that it is possible to detect at least one small galaxy ($M_{\rm vir}\approx10^{8.8}\msun$), if an active starburst takes place even at $z\approx7$ with a single pointing.}
\end{figure}

We consider two timescales for the duration of nebular emission, $\Delta t_{\rm neb}=2$ Myr and $\Delta t_{\rm neb}=5$ Myr, respectively. Compared to the typical lifetime of OB stars, of order a few 10~Myr, a shorter duration of nebular emission results in a reduction in the number of observable galaxies. This is a reasonable assumption for small galaxies, where gas clumps near newly forming stars are easily disrupted and SN feedback facilitates the escape of photons into the IGM, causing the nebular emission to be unimportant. We provide two estimates, the number of galaxies detectable in the H$\alpha$ line with a fixed mass of $10^{8.8}\msun$, and the total observable number of galaxies by choosing $M_{\rm max}=\infty$. At $z\approx7$, about one small galaxy with a mass of $M_{\rm vir}=10^{8.8}\msun$ is expected to be observed in a single field of view, using MIRI with signal-to-noise of $\rm S/N=10$ for a deep exposure of $10^6$~s. This number could reach as high as a few hundred sources, if we include galaxies more massive than $M_{\rm vir}=10^{8.8}\msun$. Unlike the small galaxy explored in this work, more massive galaxies are expected to experience continuous star formation rather than a bursty mode, giving rise to a longer duration of $\Delta t_{\rm neb}$. Therefore, the estimates of the total number of galaxies above $M_{\rm vir}\gtrsim10^{8.8}\msun$ computed here should be considered as a lower limit.

\section{Summary and Conclusions}
We have explored the assembly process and the detectability of a primordial dwarf galaxy with a virial mass of 
$M_{\rm vir}\approx10^9\msun$ at $z\gtrsim8$ by performing a cosmological, zoom-in, radiation hydrodynamic simulation,  
considering the radiative, mechanical, and chemical feedback exerted by the first and second generation of stars. 
The galaxy originates in a minihalo with a mass of $M_{\rm vir}\approx10^6\msun$ at $z\approx22$, hosting metal-free 
stars, and achieves the transition in star formation mode from Pop~III to Pop~II at $z\approx17$. At the end of the simulation 
at $z\approx8$, the galaxy contains a total stellar mass of $M_{\ast}\approx7\times10^5\msun$. Thus, small galaxies at high redshifts grow fast within $\sim 500-600$~Myr from primordial objects to complex galaxies, with properties that are similar to dwarf galaxies in the local Universe.  

In order to investigate whether such low-mass dwarf galaxy ($M_{\rm vir}\approx10^9\msun$ at $z\gtrsim8$) can be observed with upcoming frontier telescopes such as the {\sc JWST}, we post-process the simulated galaxy using the spectral energy distribution generation code {\sc SUNRISE}, and also separately compute the nebular emission directly from the gas properties in the simulation. We have found that the galaxy is too faint to be detected in the UV continuum throughout the simulated cosmic time $z\gtrsim8$, while the recombination luminosity, in particular for the H$\alpha$ line, might be within reach of the detection limit for MIRI onboard the {\sc JWST}, if an active starburst takes place. The simulated galaxy experiences a starburst 
at $z\approx9$ with a star formation rate of $\dot{M}_{\ast}\approx0.1\msun\rm yr^{-1}$, producing a recombination luminosity of $L_{\rm H\alpha}\approx2\times10^{40}\mbox{\, erg} \mbox{\,s}^{-1}$, which is comparable to the MIRI detection limit for a deep exposure of $10^6$~s with a signal-to-noise of $\rm S/N = 10$.

Whether the nebular emission from the first dwarf galaxies can be detected depends on the gas properties in the vicinity of star forming regions. Nebular emission is only prominent when the gas is in a high-density phase, which corresponds to a time period immediately after the starburst. Unlike more massive galaxies, which tend to sustain continuous star formation, we find that the simulated first galaxy is experiencing a bursty formation mode. This is due to the vulnerability of low-mass galaxies to stellar feedback, specifically the photoionization heating and supernova explosions that vigorously disrupt the gas near star forming regions. As a consequence, the time period over which nebular emission is sufficiently strong for detection, is limited to $\lesssim 3$~Myr after the starburst, lowering the probability of detection. Considering the duration of nebular emission, we derive the number of galaxies with a mass of $M_{\rm vir}\approx10^{9}\msun$ above the H$\alpha$ emission line detection threshold at a given redshift. We suggest that about one such galaxy can be observed with MIRI in a single pointing. Furthermore, the number of detectable sources may be significantly boosted, if such primordial galaxies are strongly lensed with a magnification of $\mu \gtrsim 10$ (e.g. \citealp{Zackrisson2012}).

Indeed, it is extremely difficult to observe such low-mass galaxies at high-$z$, even with the JWST. If, however, such galaxies experience intense starbursts ($\dot{M}_{\ast}\gtrsim0.1\msun\yr^{-1}$), emitting strong $\rm H\alpha$ and other nebular lines such as $\rm [O~III]$ 5007 {\AA}, they may be detectable in multiple lines with the JWST in ultra-deep exposures ($10^6$~s), or if the galaxies are gravitationally lensed. A complementary technique, if multiple line detections are not feasible, may be to photometrically identify the break between near-IR and mid-IR bands, with no detected flux in the former and a detection for the latter due to the contribution from H$\alpha$. Given the properties of the galaxy simulated here ($M_{\ast}<10^6\msun$, $\dot{M}_{\ast}\lesssim0.1\msun$\,yr$^{-1}$ at $z\gtrsim8$), this system may thus constitute the very limit of detectability with the JWST. It is exciting that observations will soon begin to reach the most pristine star forming systems in cosmic history, thus testing our model for cosmological structure formation.

\section*{acknowledgements}
We thank the anonymous referee for constructive and insightful comments that improved the clarity of our paper. We are grateful to Volker Springel, Joop Schaye, and Claudio Dalla
Vecchia for letting us use their versions of \textsc{gadget} and their data
visualization and analysis tools. M.~J. is supported by a National 
Research Foundation of Korea (NRF) grant (NRF-2018R1C1B6004304), funded by the Korean government (MSIT). V.~B.\ acknowledges support from NSF grant AST-1413501. The authors acknowledge the Texas Advanced Computing Center (TACC) at The University of Texas at Austin for providing HPC resources under XSEDE allocation TG-AST120024.
\footnotesize{

\bibliography{myrefs2}{}
\bibliographystyle{mnras} 








\bsp	
\label{lastpage}
\end{document}
